The work in this report has been carried out as part of a commercial project for a published game title named "*Fallen God: Escape Underworld*", in which all the reported results are fully integrated, something that can be directly observed by the relevant videos in the following URLs:

Apple Store (iTunes) product page:

https://itunes.apple.com/kh/app/ankh-the-light-prevails/id1086373626?mt=8

Facebook product page:

https://www.facebook.com/fallen.god.escape.underworld/

# There is more to PCG than Meets the Eye: NPC AI, Dynamic Camera, PVS and Lightmaps

*Anthony Savidis*,
Computer Science Department, University of Crete and
Institute of Computer Science, FORTH
`as@ics.forth.gr`

**Abstract.** Procedural content generation (PCG) concerns all sorts of algorithms and tools which automatically produce game content, without requiring manual authoring by game artists. Besides generating complex static meshes, the PCG core usually encompasses geometrical information about the game world that can be useful in supporting other critical subsystems of the game engine. We discuss our experience from the development of the iOS game title named 'Fallen God: Escape Underworld', and show how our PCG produced extra metadata regarding the game world, in particular: (i) an annotated dungeon graph to support path finding for NPC AI to attack or avoid the player (working for bipeds, birds, insects and serpents); (ii) a quantized voxel space to allow discrete A* for the dynamic camera system to work in the continuous 3d space; (iii) dungeon portals to support a dynamic PVS; and (iv) procedural ambient occlusion and tessellation of a separate set of simplified meshes to support very-fast and high-quality light mapping.

## 1 Introduction

Today, there are numerous applications of procedural content generation (PCG) into games, all effectively relying on the idea that some kind of an algorithm is assigned the task to regenerate what usually humans will have to manually design. The story of the technique itself goes well back in time, with Rogue, a 2d dungeon crawler game from the 80s, attributed as one of the earlier games to feature PCG for dungeon generation during gameplay. More recent and widely popular games are Diablo III and Far Cry 2, with emphasis on visual geometry (polygonal meshes). Overall, there is a clear trend to adopt PCG in production-level games, and in some cases to provide the PCG engine with appropriate configuration editors to the players themselves.

The problems inherent in content automation are well known, mainly related to quality, diversity and believability of the generated artifacts [4]. The reasons justifying this ever increasing trend vary, ranging from the constantly increasing capacity of GPUs than can easily handle the rendering of massive procedural content, to the extraordinary demands for huge amount of content and detail in modern games, which, when manually crafted, can in many situations overflow the budget of the average game developer. In this paper, our emphasis is shifted from the traditional PCG core mechanics and implementation details to the actual benefits it can bring to the rest of the game engine subsystems. More specifically, any PCG tool is expected to necessarily involve high-level information regarding the game world or the terrain structure, and to output some extra data, besides visual content, that is used by the core game logic. The type of such data, that we will also mention as generated meta-data since it is extra information for the automatic content, varies from mesh indexing, to level of detail mapping, to occlusion culling purposes. We will discuss how such generated meta-data played a key role in supporting important game features across various subsystems, showing that there is more to PCG than graphical content, visual appeal, geometry computation, and artwork style. The latter is exposed as our initial experience in a real-life production-level game named 'Fallen God: Escape Underworld', see official App Preview video[1]).

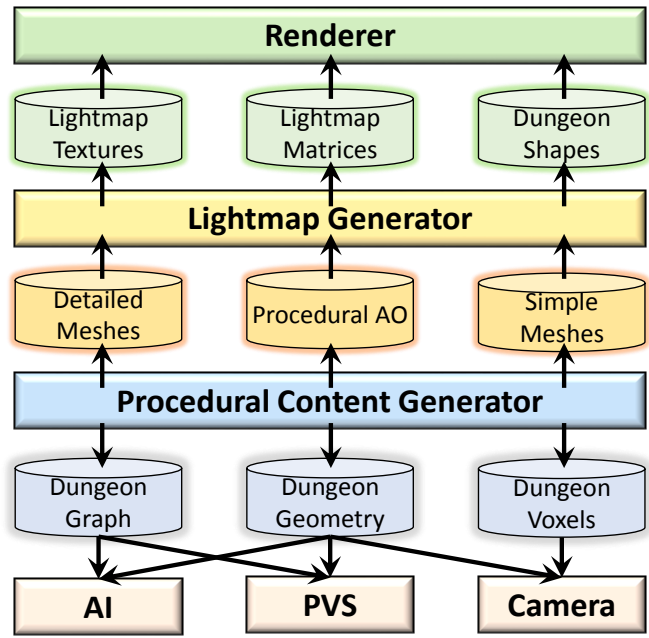

**Fig. 1.** The central role of the PCG component producing extra information besides meshes (additional geometry, kind of metadata) that feeds other critical game subsystems.

### 1.1 Contribution

We discuss how we exploited the PCG module to automatically produce detailed meta-data in a way supporting four independent subsystems of the game engine (see Fig. 1), in particular: (i) the NPC motion-control AI for various creatures including bipeds, arthropods, bats and a serpent composed of independently coordinated skeletal parts; (ii) dynamic (third person, tracing) camera system, relying on optimal positioning behind the player, and using an A* path finding algorithm for camera repositioning working on a procedurally quantized terrain space; (iii) dynamic Potentially-Visible Set (PVS) with portals; and (iv) lightmap generation with aggressive static batching of world meshes relying on procedurally generated and indexed point lights within dungeons, and on mesh adjacency and grouping information.

All the latter are supported by geometry and connectivity information produced by the PCG module in additional to polygonal meshes. Moreover, besides the typical query functionality to access such information, extra utilities methods were provided, utilized by various subsystems. For instance, the tagging of elements to dungeons served originally the PVS subsystem for culling purposes of static geometry, and the lightmap generator for better grouping of the lightmapped meshes together. However, the exact same functionality worked just fine for tagging dynamic elements, such as NPCs and special effect items, as part of the PVS for dynamic geometry. Thus, not only the querying, but almost all edit-ing functions, of the PCG metadata were put together as a core backbone to support the previously mentioned subsystems.

---

[1] https://www.facebook.com/fallen.god.escape.underworld/videos/1192517717459991/

Overall, the primary contribution of our paper is a full-scale real-life application and experience reporting on the way PCG metadata, in a production-level game, constituted the backbone for a set of critical game subsystems. There is also earlier work in PCG techniques related to game content apart of polygon generation, ranging from plot sequence to terrains recursive structures. In this context, we consider that PCG should go beyond visual realism and artwork automation, and emphasize the production and runtime manipulation of more logical information regarding the overall game logic and the way it is internally exploited by the various game engine subsystems. One example of the dungeon graph visualization, part of the meta-data, during gameplay (the text labels and plane normals are disabled for clarity), is shown under Fig. 2.

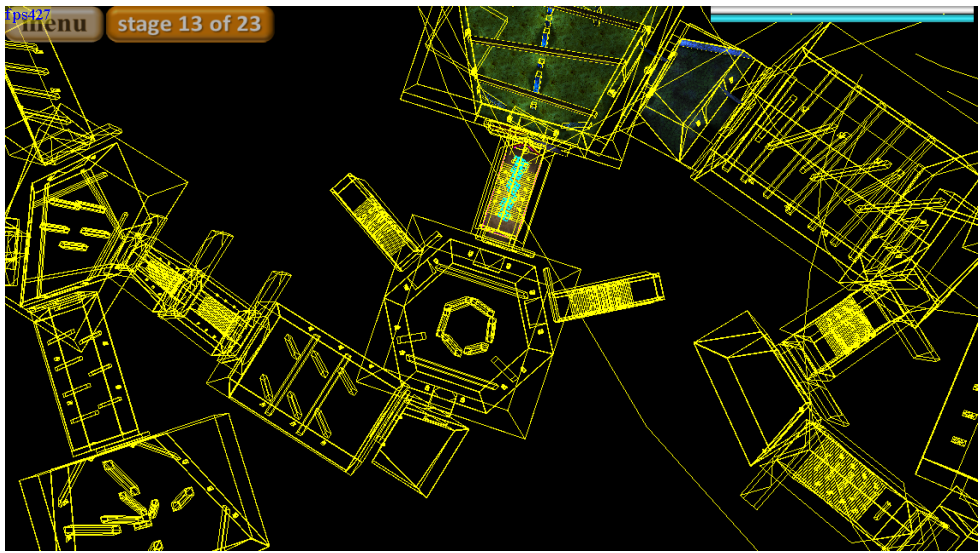

**Fig. 2.** Dungeon connectivity graph annotated with high-level geometric information (e.g. planes, volumes, angles, heights) for all contained elements (sides, floors, ceiling, columns, gates, sliding doors, torches, stair steps and stair bases).

## 2     Motion Control AI for NPCs

The motion control AI is a key ingredient of action games as it regulates the way characters can move and behave in the game arena, with viability of a motion request also concerning the player. For NPCs the primary goal is to control them in a way better and more naturally challenging the player. For instance, once expects a different motion behavior between bats and mummies, regarding reaction, speed, direction and coordinated group behavior. For players, the primary target is to assess motion viability against the terrain structure, and make corrections such as walking correctly sideways to walls, avoiding timely columns without causing motion loops, and in our game effectively controlling flight mode, by computing smooth and plausible landing paths in real-time. Earlier work [12] shows that precomputation of NPC motion paths is possible, depending on game genre, even in large environments, turning motion control to data lookup. In our game, due to NPC AI, such a method was impractical.

Overall, the motion AI mainly takes into account (dynamically) the following information: (i) current player position and direction; (ii) current position and direction of all NPCs; and (iii) the geometric space in the current focus of action and attention. In particular, the part regarding the geometric space is very crucial and is thoroughly output by the PCG engine during production time, and includes elements such as: volumes (combining spheres and oriented bounding boxes), side planes, polyhedral objects (for columns), dungeon height maps, direction of linked tunnels and stairs from dungeon centers, gates and doors central (cross) and side planes, and dungeon connectivity information. Effectively, for every character in the system we track its current dungeon, making dungeons the primary dynamic containers on which all static or dynamic game objects are finally attached.

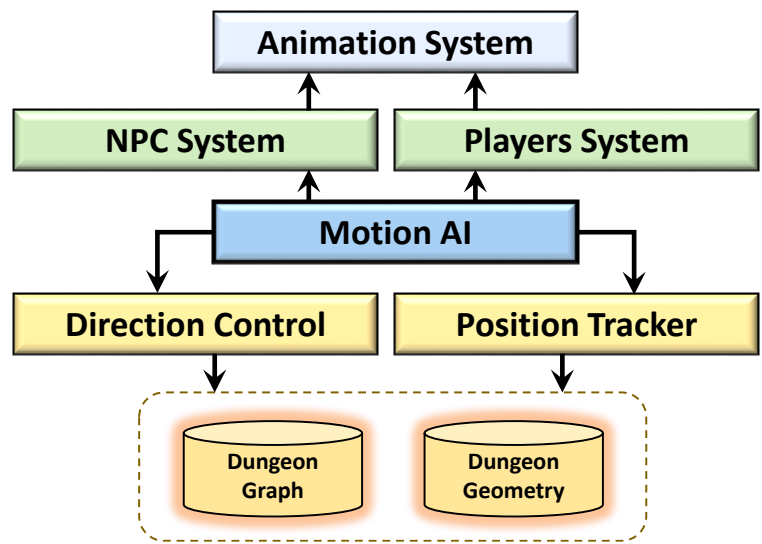

**Fig. 3.** The motion control AI for NPCs and its contact site with other systems, showing how it heavily relies on the dungeon meta-data produced by the PCG engine.

The general architecture of the motion-control system is depicted under Fig. 3, while the detailed structure of the actual component implementations is shown under Fig. 4. Our emphasis is put on setting the *PCG Data Access* module as a first-class runtime component of the actual game engine. In fact, this way the component is open-ended with respect to the geometric information it can incorporate, something that we practically exploited during the development lifecycle of the title itself. More specifically, the scorpion NPC, as well as the Horus fly feature, have introduced in the game after version 2.5. All extra required geometric information from the dungeons to carry out the required motion tests was included in the PCG meta-data and related data-access with no additional amendments to the rest of the engine modules.

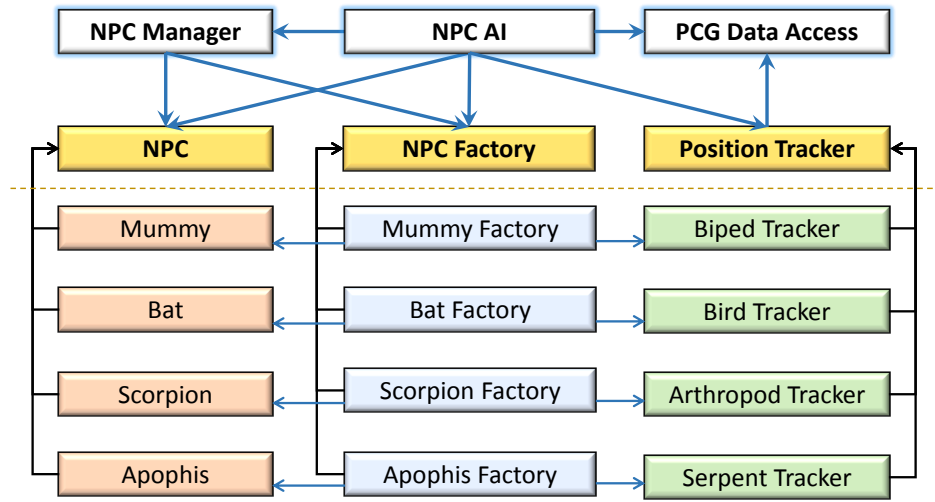

**Fig. 4.** The class relationships for the implementation of the NPC AI and the central role of the PCG meta-data being part of the high-level engine modules (i.e., above the horizontal dotted line, denoting the invariant game engine modules - those below are subject to extensibility).

The importance of geometrical meta-data is very critical. It allows for more performance efficient tracking and motion control implementations, since all related geometry is precomputed, including tracking and pre-caching of motion singularities. For instance, invalid passages, regions where player does not have enough room to fly, locations where turns should be constrained to specific directions to avoid facing the wall, zones where the column formations are very dense and Apophis cannot pass without visually-evident intersections, are, just to name a few, typical cases where the PCG produces tables to precondition various direction update choices.

One of the most impressive and well-accepted features of the motion control for NPCs in Fallen God concerns the precise, timely and fast flip turns of the bats. In fact, it is also shown in the first 2 seconds of the official App Preview of the title. An example of such an attack sequence during gameplay is explained under Fig. 5.

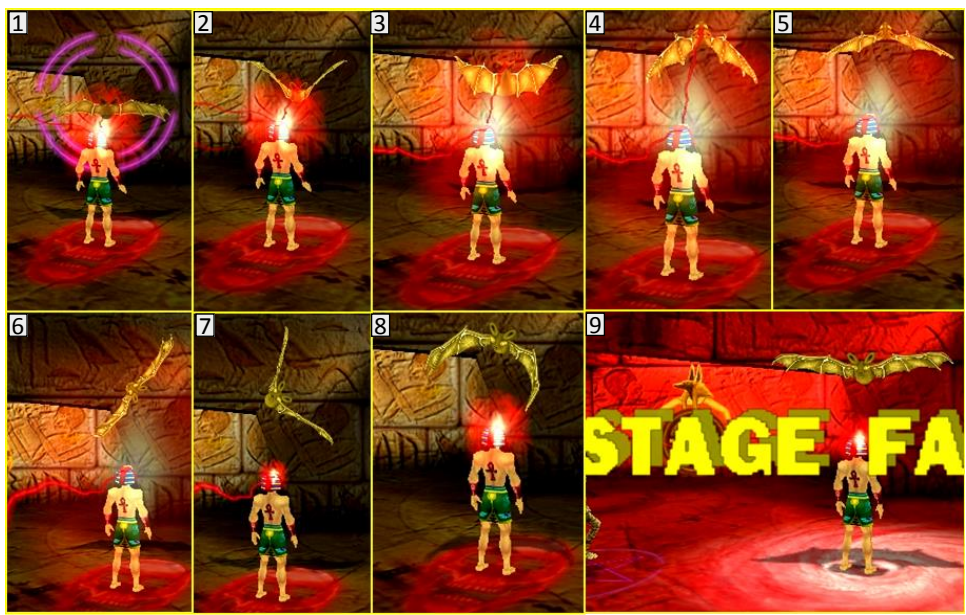

**Fig. 5.** Bat attack motion-control sequence: 1, 2: approaching player from behind; 3, 4, 5: the front wall plane is too close, while the left / right wall planes give no room for left / right turns, thus a flip turn maneuver is chosen; 6, 7, 8: the bat turns and balances, computing forward space to opposite wall plane; 9: it launches an attack directly and eliminates the player.

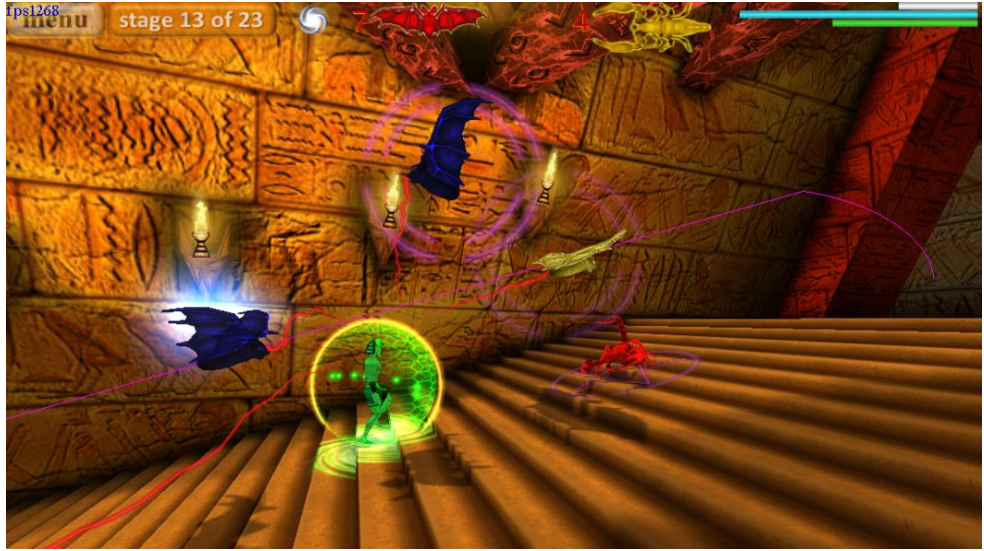

**Fig. 6.** The motion control differences between different character types: (i) bats have to avoid walls and stay at a correct altitude while performing fast maneuvers and precise flip turns; (ii) scorpions should crawl on stair steps, apply balanced turns on inclined planes, while avoid each other, and attack (with the sting) the player if in close range; and (iii) the player should be aligned perfectly to stairs steps, even when landing on them.

The motion-control requirements of the various characters vary a lot, and require different and detailed geometric information (see Fig. 6). For instance, scorpions can crawl on stair steps, requiring information regarding step planes, base planes (start / end of stairs), transition planes (when shifting from base to steps and vice versa), and wall planes. Additionally, as gates are smaller and lower to dungeons, we also precompute the transition path for bats, from the dungeon to the gate and vice versa, while applying a 90 degrees angle side turn during altitude change.

In fact, the most complicated motion control concerned Apophis (see Fig. 7), the eternal enemy of Horus, which, being a giant serpent, moves in-between the column formations. For the visuals, due to the requirement for running on low-end iOS devices (starting from iPhone 4, with Open GL ES 2.0), we had to drop GPU skinning of the mesh thus a vertebrate structure of separate components was adopted. In this case, the sinusoid motion of the serpent involved continuous checks with walls and columns, to compute maneuvering and direction changes; interestingly, we reused the exact PCG data that were required for controlling bat motion.

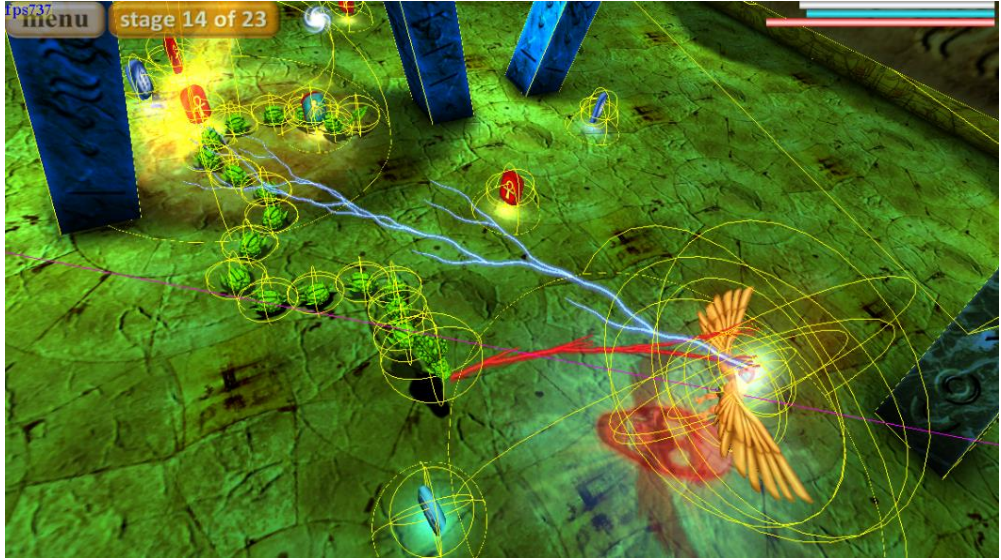

**Fig. 7.** The motion control of Apophis is distinct and more complicated implementing the typical coordinated sinusoid skeletal locomotion of serpents; it heavily relies on PCG generated world geometry, in particular minimum distances to columns and walls, column voxels and precomputed plausible direction changes, in order to ensure there is enough room for effecting maneuvering either towards or away the player, or when avoiding column formations.

## 3    Dynamic Camera System

Dynamic camera systems are also known as real-time cameras [8], with numerous techniques currently existing for either first or third person perspectives [20]. The camera system of the game title is a dynamic third-person tracing camera and it had to obey one invariant design rule that was set very early: *should never require turning either transparent or completely hidden any of the visible geometry due to camera repositioning*. In other words, the camera should move around geometry in real time and with a high degree of visual fidelity, requiring minimum distance transitions. Additionally, since the camera can be dynamically configured by the player, in terms of altitude, distance to player, and targeting (down) angle, then all computations should be necessarily applied in real-time within 3d space.

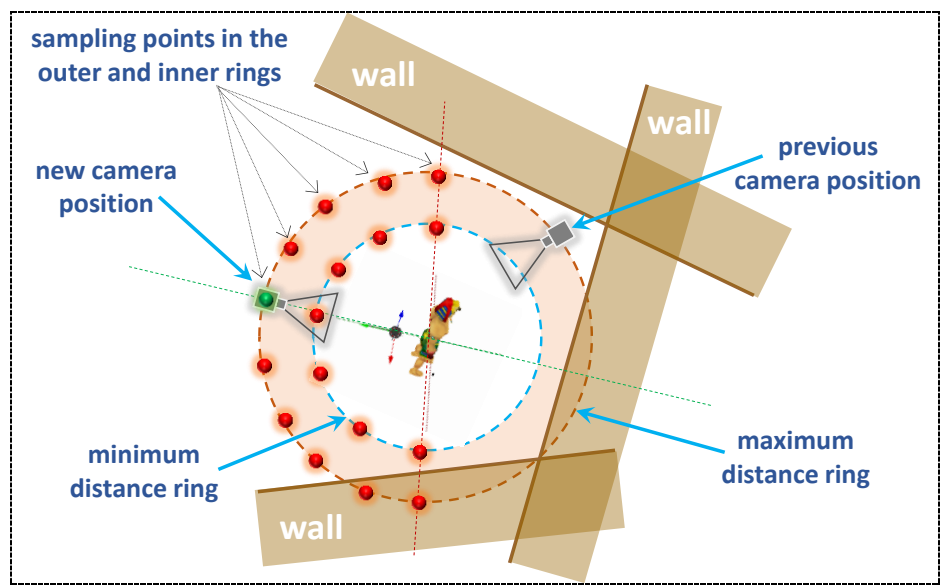

**Fig. 8**. First phase of camera repositioning, with computation of a plausible best-fit new location reacting to player position change; applies dense sampling in a number of zones around the player, between an outer and inner ring (min / max only shown here); such calculations apply volume / plane intersection tests amongst the camera, walls, columns, torches and gates.

The camera system of the game implements a three stage update process: (i) an appropriate camera position is initially computed, which, as shown under Fig. 8, involves intensive intersection tests with dungeon volumes and planes during camera position sampling; (ii) a path finding algorithm is run to compute the transition from the previous to the new position, while applying smoothening for steep turns; and (iii) camera animation is launched to perform the transition path, with an adaptive motion speed according to the camera-player distance – an update can be retriggered during this stage if the player gets too close / far to the camera. From those three phases, it is the second one which requires to compute an appropriate transition path from the previous to the new camera position.

Apparently, there is no guarantee that a straight path is a good solution as the distinctive column formations of the game may well, and most likely will during gameplay, obstruct the straight line between the two camera positions, causing the camera to pass through visible geometry, something unacceptable from the design requirements. When path finding is involved in games it is common to adopt a popular algorithm like A* since it offers real-time performance with guaranteed optimal results [5]. However, the problem is that it only works in grid spaces, hence, one has to quantize the 3d universe to ensure it is applicable. Also, since the quantized space is still 3d, with height, we need to organize it appropriately into layered grids and apply A* on the grid corresponding to the particular camera altitude. Typically, every grid element has a flag indicating whether it is empty space or solid (i.e., intersects with environment meshes). The choice of the voxel size in the quantization affects the dimension and density of the grid, and is proportional to the performance of A*, while inversely proportional to the precision and smoothness of the resulting path. In general, a reasonably dense grid allows fast and precise A*, offering more solutions for positioning, while enabling paths that once smoothened work very well with animation. We discuss how terrain quantization was part of the overall PCG process.

### 3.1     Terrain Quantization

To support 3d path finding for the camera, the terrain quantization of the dungeon spaces has been entirely implemented as part of the PCG process, which produced the dungeon meshes themselves, supporting per dungeon configurable voxel sizes and number of grid layers. The latter reflected not only the variations on dungeon structures, and in particular the column formations, but also the allowed camera altitude variations throughout the entire game. Under Fig. 9, the quantization is shown within two dungeons of the game having relatively complex procedurally-generated column groups. In many cases, and in particular while the player is flying, the camera performs very dynamic and immersive updates, following the player while maintaining min / max distances (for the layer, a spring-based approach was also implemented, but does not involve PCG data). The need for quantization of the game space has been identified earlier in [6], and was implemented as part of a post-processing phases on game meshes to realize a *transition-planning camera system* (Patent US8576235).

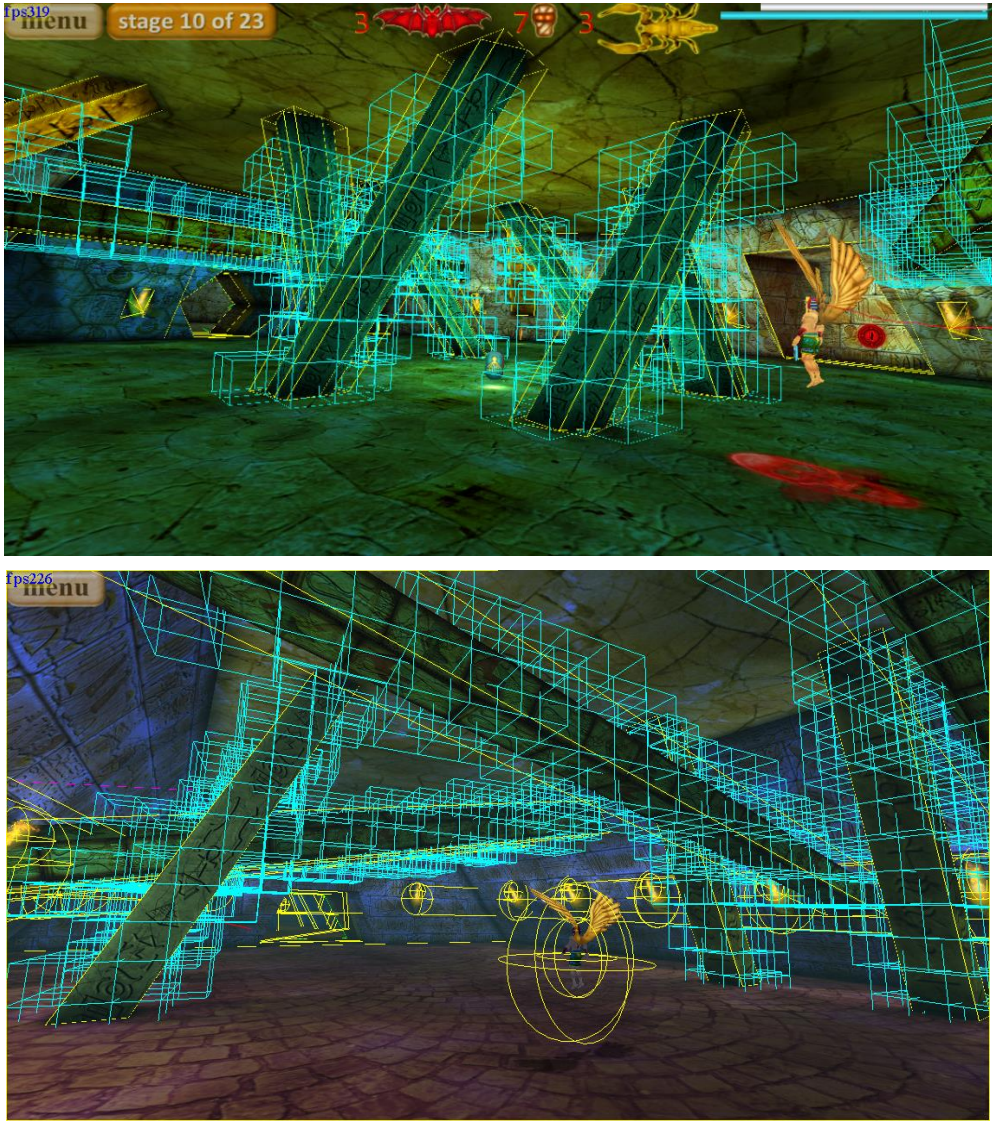

**Fig. 9.** Voxel space for the complex column formations inside dungeons, enabling to quantize the camera motion space and allow fast and precise A* path finding in a discrete space; used for computing the repositioning path of the camera in real-time without making columns temporarily transparent.

## 4 Dynamic PVS with Portals

Potentially Visible Set (PVS) [3] is a technique commonly implemented in video games, originally introduced by the Quake game engine, and supports hidden object removal (occlusion culling), according to which a set of potentially visible elements is statically calculated (during development time, always an estimate) and is associated / indexed to every possible terrain region. Then, it is appropriately queried and used during gameplay to render only what is currently visible from the active camera configuration. Its purpose is to limit down the number and size of rendering batches as much as possible.

The dynamic version of the PVS algorithm applies a similar method by heavily relying on meta-data regarding the terrain geometry, besides mere use of polygonal meshes. The most common dynamic PVS method is portal-based, and was initially introduced in [10], working for in-doors or closed space areas, with such areas visible only through doors, gates, windows, or similar geometrical holes known as portals. There are variations as in [1], while in general its performance is dependent on the precision of the occlusion culling calculations inherent in the portal volumes: the more precise the portals are, the smaller the probability for visibility false positives. Dynamic caching can be combined with portal-based PVS, for both portal volumes and visible elements per portal, while if memory requirements are sensitive, a generic dropping policy such as LRU (least recently used) can be used, together with game-specific policies such as dropping abandoned places (i.e. terrain areas that the game plot will never allow to return).

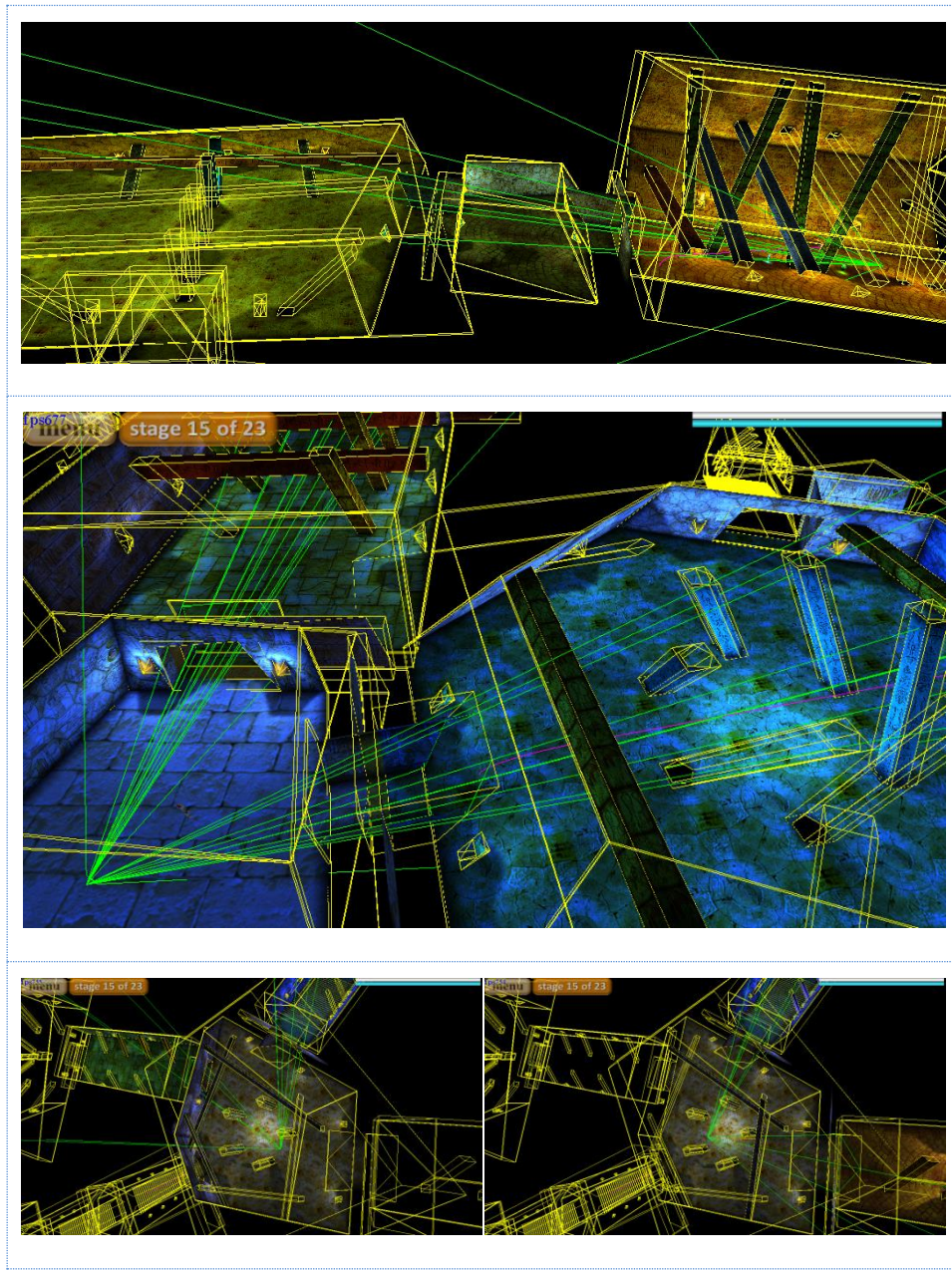

**Fig. 10.** Dynamic casting of portals from the current camera position (start point of the green line beams, usually behind the player) by tracing dungeon connectivity and visibility of elements through gates and doors; even small changes on camera (player orientation) cause PVS reevaluation that works perfectly in real-time thanks to the large part of static information (volumes / planes shown with yellow outlines) generated by the PCG engine.

The type of PCG data utilized by the portal-based PVS algorithm are firstly the gate and door geometry together with the grouping of meshes into dungeons, secondly, the bounding volumes and planes that are also produced by the PCG engine, and finally the dungeon graph itself. The portal algorithm recursively casts portals from the current camera position to the gates and doors currently visible, until no further visible gates or door can be traced.

Also, the particular structure of the terrain in Fallen God challenges any portal-based algorithm due: (i) intense connectivity of dungeons with the multiple levels of chambers connected through stairs; (ii) the stylistic choice to avoid orthogonal angles everywhere, except from stairs steps, something making typical optimization calculations met in common human buildings overly impractical (there are polygonal dungeons, distorted gate shapes, scaling over distance tunnel sections, and inclined ceilings); and (iii) the design decision that *no abandoned places exists in the game*. The latter implied that as the player progresses stages, the visited dungeons remain visible and can be visited, thus potentially increasing the complexity of the PVS.

In Fig. 10, we show snapshots of the PVS in action, for various dungeon configurations, ranging from successive tunnels with intermediate rooms, to cyclic structures of tunnels around a large chamber; the PCG geometrical meta-data are also rendered in yellow. Its implementation relies both the statically and dynamically computed elements that are associated to dungeons, as depicted in Fig. 11. The respective PVS method is detailed and aggressive, dropping everything not viewable by the portal. To do this, it relies on the bookkeeping of all elements by every dungeon, with the most visually intensive polygons already statically produced, linked and indexed by the PCG module.

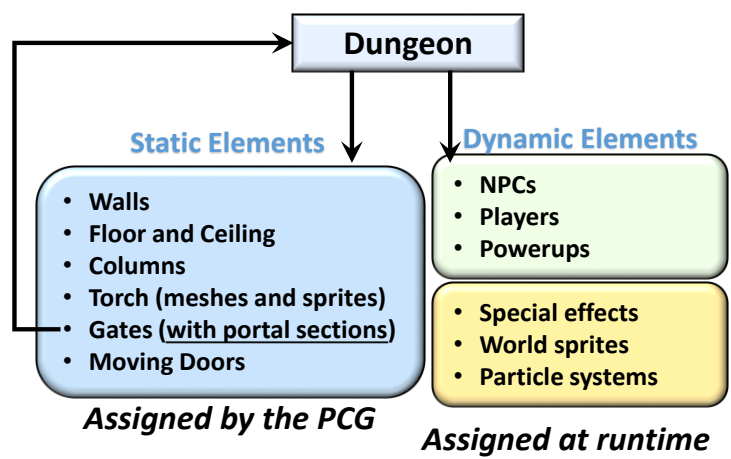

**Fig. 11**. The larger part of the visual elements in a dungeon are statically associated and grouped by the PCG module, including their respective bounding volumes and the portal sections of games, used during portal creation and volume intersection tests.

## 5   Lightmap Generation

A lightmap is a texture whose pixels sample the illumination received from all light sources at a rectangular area of a world surface. Effectively, a set of light maps encompasses the overall illumination received by all surfaces of a word map. Usually, high-resolution lightmap atlases are produced (see Fig. 12), combing the individual lightmaps for a set of surfaces. It is considered a pure static lighting method since no illumination computations are involved at runtime, and is popular production-time preprocessing method as everything related to lighting is computed during development time. The generation and deployment of lightmaps is a complex topic with numerous methods and variations, some of which denote dynamic computation and caching of lightmaps.

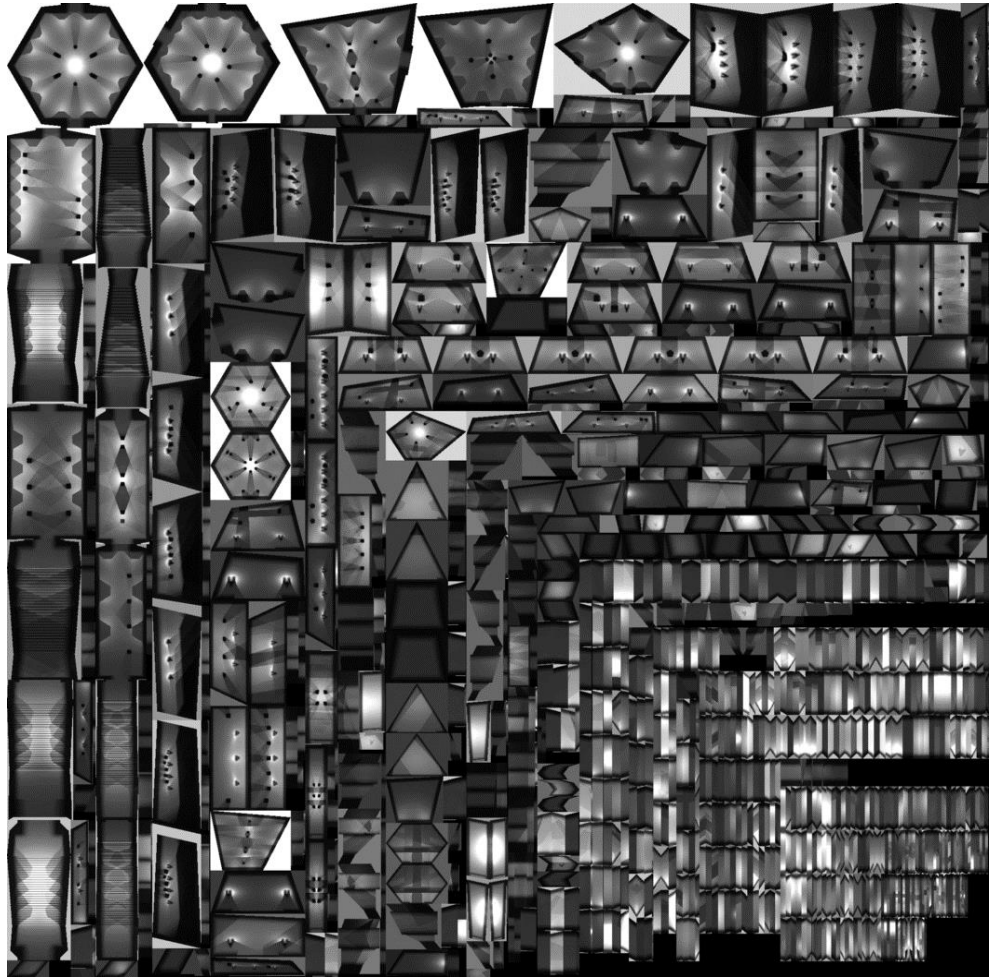

**Fig. 12.** Lightmap atlas generated for the entire game utilizing the dungeon graph, geometry and detailed side planes produced by the PCG module.

The way lightmapping links to PCG is explained by the type of information required to extract and group surfaces together into the same lightmap. The lightmap production process is depicted under Fig. 13 process. We explain its details in linkage to PCG meta-data. In particular, our lightmap production had to address the following requirements: (i) surfaces should be reasonably and logically proximate to each other in the world map to avoid bringing a lightmap involving objects that are not rendered together (i.e. far to each other), that is, optimal batching should be preserved; (ii) scaling of lightmaps should rely on the way the surface is actually viewed during gameplay, meaning we use larger lightmaps for objects viewed close and for some time compared to objects where focus of attention is sporadic and incidental.

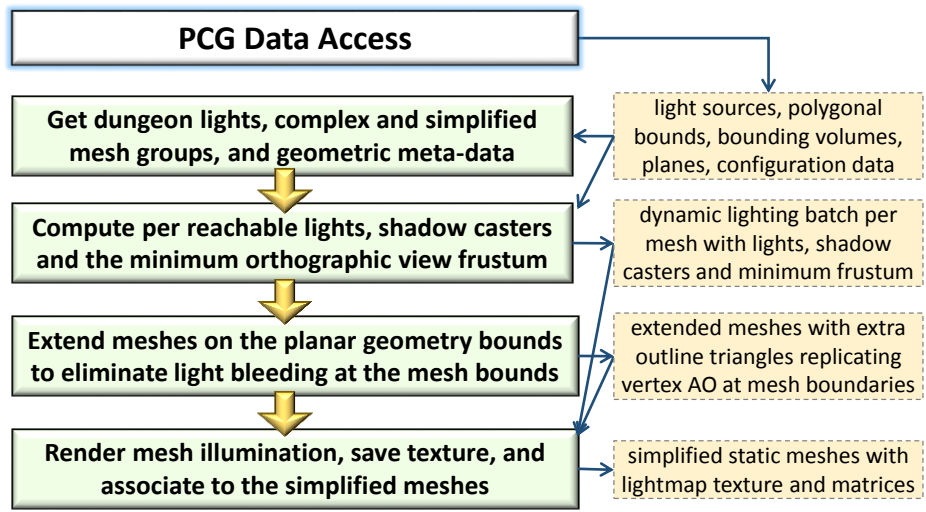

**Fig. 13.** Overview of the lightmap generation process and its strong coupling and dependency on the geometry meta-data of the PCG process.

Now, both previous requirements directly concern semantic information regarding the surfaces themselves, something that is usually resolved through developer intervention (configurations and directives) for correct grouping, commonly as part of the lightmap tool-chain in 3d popular modelling tools. But when everything is done with a PCG tool, without such configuration, the lightmapping cannot produce optimal results. In our case, the solution relied directly on the PCG engine output:

- All meshes have category identifiers (option of the PCG engine), such as "wall", "ceiling", "column", "floor", etc., so that they can be grouped in lightmap generator configuration
- All meshes are grouped (tagged) into dungeons, the latter also named by the developer, again serving grouping configuration purposes
- Mesh planes and bounding volumes are already produced by the PCG, as discussed earlier, thus helping the lightmap generator create planar groups
- Custom lightmap scaling is allowed for an entire dungeon, affecting all its elements, or for individual named meshes
- Grouping of lightmaps into various atlases follows the connectivity of dungeons: every dungeon shares the same lightmap either of its previous or next dungeon in the logical game progress
- Similarly, the light sources involved in the computation of a lightmap are those from the current and neighbor dungeons – again dungeon volumes are involved to seek affecting point lights

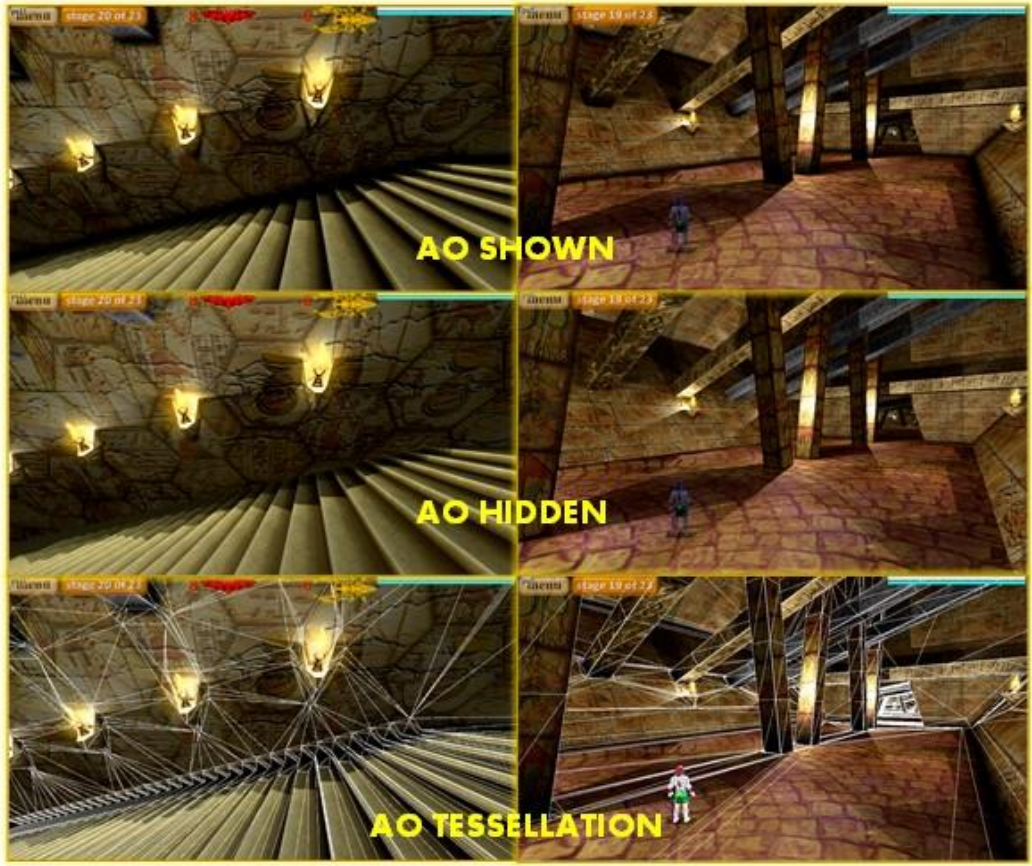

**Fig. 14.** Custom procedural tessellation applied at the intersections between walls, columns and stone torches, to exaggerate and amplify the ambient occlusion effect and make to make it more intense and apparent as part the light map generation (shadows are omitted in the snapshot to make AO the more evident).

In general, we very quickly observed a strong dependency of the lightmap generator to the actual PCG engine meta-data. In fact, one may argue that lightmapping is essentially another PCG method, where the procedurally produced content is the precomputed illumination in the form of textures. During the implementation of the light mapping tool we initially decided to combine it with dynamic ambient occlusion (AO), in particular with SSAO (screen-based ambient occlusion), originally introduced in the game Crysis by Crytek [11] with improvements like [13]. The performance we accomplished in an iPhone 4 was not what we actually expected, while trying sampling reduction caused unacceptable visual artifacts. Thus we decided to adopt the traditional mesh-based ambient occlusion, while bake it together with the dynamic lights during lightmapping. But it is known that vertex-based AO is mesh-dependent, more specifically is tessellation dependent, meaning it is very hard to obtain high quality results with low polygon tessellation. Additionally, we wanted to give an exaggerated artistic tone on the ambient occlusion at wall intersections, in order to emphasize the mystic atmosphere inside the underworld dungeons. To support this, it was quickly realized that a custom tessellation approach for such style of ambient occlusion was required as part of the PCG mechanics.

Consequently, we ended up with a *procedural ambient occlusion* method, which resulted into more complicated meshes (i.e. more triangles were needed). To avoid using such meshes in deployment time, and thus increase excessively the vertex shader performance overhead, we choose to produce very simple meshes for planar surfaces that were to be deployed together with lightmaps. The latter justifies how Fallen God is capable to retain a low-bound of 40 FPS rate even in an old iPhone 4 device. Under Fig. 14, snapshots showing the procedural AO being on / off, and its tessellation details are provided; the latter concern wall bounds, stair steps, outlines of wall mounted torch bases, and the intersections of columns with walls, floor and ceiling. In these snapshots, the mesh coloring used in the title is disabled for clarity of the images. Overall, we observed a very strong interplay between the lightmap engine and the PCG engine, in particular the former strongly relies on the latter, in a way that made us directly draw the conclusion that we should also treat lightmapping as a sub-category of the PCG universe.

# 6 Related work

Procedural content generation for games is a huge topic with many tools, domains and disciplines, ranging from graphical detail and realism, to automatic generation of the design aspects using machine learning, rules or even evolutionary techniques. There is a rapid proliferation of the games embodying PCG in many respects, however, it is a common practice for published games to keep the technical details closed. In this sense, although many games introduce severe innovative PCG features, for industrial reasons they constitute proprietary information. Recent games such as *Astroneer* [16], apply PCG to produce large-scale worlds (planets), while providing to the player a *terraforming* capability, essentially offering features of the PCG subsystems to the player in friendly and easy manner, turning PCG to a direct game mechanic.

Another worth mentioning game is *Elite: Dangerous* [17], which combines PCG with large-data (astronomic, regarding star systems, galaxies and constellations) to procedurally produce a huge number of new star systems by combining data from existing ones. In a similar way, *No Mans's Sky* [18], is another great game featuring planetary exploration and survival challenges, while offering procedurally-generated flora and fauna of planets. Interestingly, this game can produce far more planets that any human player could actually visit, even with the fastest possible exploration pace, in his entire lifetime. Clearly, compared to our discussed title, these games provide massive content and are targeted to high-end consoles, rather than to average mobile devices. Technically, their emphasis is shifted in massive visually realistic content, since the scale of the world and its artistic impression are part of the first-class game elements. All previous games realize some form of guided PCG, from existing structures and instance, to automatically produce new spaces reflecting connectivity semantics inherent in the original supplied models and prototypes.

Besides the work done in the context of industrial games, there is research work focusing on aspects related to user experience [7, 9], playability, believability and semantic coherence of the procedurally generated game spaces. The work in [1] emphasizes the automatic generation of the entire dungeon space without guidance or design-provided connective, size or style directives. It should be mentioned that in our title, we adopted guided PCG using style and connectivity rules for the dungeons, main reason we could not automate entirely everything was the need to tune plot escalation and atmospheric details, something not possible without precise design configuration in the PCG process.

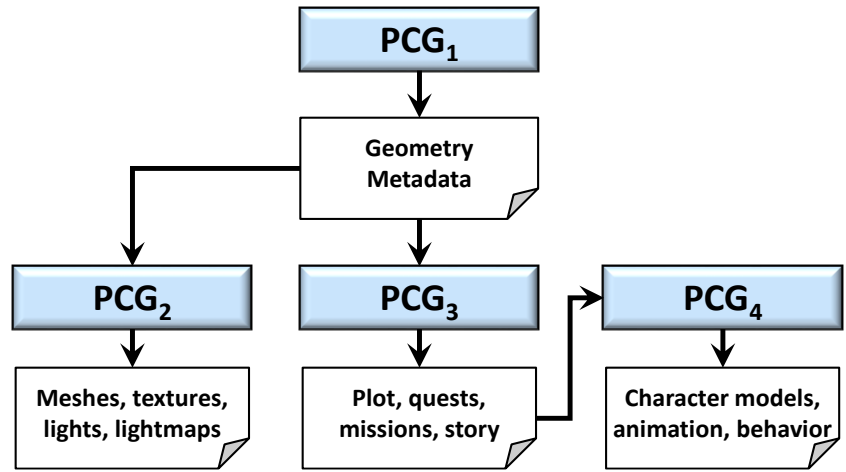

**Fig. 15.** Illustrating the idea of multi-layered pipelined PCG, combining alternative engines, by standardizing the intermediately produced content and data; the processing levels mentioned are only indicative.

In general, although our emphasis in this paper is still on the generative nature of a PCG engine, our focus shifts from the content itself to the actual meta-data that can be produced linking to this content. In particular, during the generation process, the PCG tool creates and keeps a lot of different types of geometric elements, we call them geometric meta-data, which can make the life of a game developer much easier by supporting various demanding game techniques, like those discussed in the context of the paper. In fact, it can help create multi-level PCG techniques, where each PCG module produces output for another PCG layer or third-party engine in the tool-chain, as outlined under Fig. 15. For instance, the work in [14, 15], regarding options for automatic story, quest and planning generation, might well concern PCG components mounted at PCG3 level of the previous figure. Similarly, the astonishing procedural NPC generation core of No Man's Sky [19] could fit into PCG PC4 level.

## 7    Conclusions

In this paper we focused on the PCG process and the way it can produce geometric meta-data besides the typical terrain maps, polygonal meshes and procedural textures. Such meta-data can effectively support the implementation of various techniques within critical game sub systems. Our argument is based on the inherent complexity of the PCG engine and the information it actually possesses or produces anyway, during processing time, so as to automatically generate the resulting content. In other words, it is a common PCG practice that, before the computation of the detailed visual elements, internally, some high-level world representation and geometry is always produced. In our context, departing from real-life practice in the production of a commercial game title, we emphasize the modeling, packaging, delivery and access of such meta-data as first-class game content. In the context of our game, it supported computations involved in: NPC motion-control AI, dynamic camera system, potentially visible set with portals, and lightmap generation.

Overall, we consider that this work is directly complementary to the core PCG techniques and methods, showing that the role of PCG engines should be upgraded to concern extra categories of game content, becoming a key component crossing and bridging various development disciplines, besides visual content and art.